\documentclass{moriond}

\def\Journal#1#2#3#4#5{{#1} {\bf #2} (#3) #4 [#5]}

\def\NPB{{\em Nucl. Phys.} B}
\def\PLB{{\em Phys. Lett.}  B}

\def\PRD{{\em Phys. Rev.} D}

\def\JHEP{\em JHEP}
\def\EPJC{{\em Eur. Phys. J.} C}
\def\AFB{$A_{\mathrm{FB}}$ }
\def\AW{$A_{W}$ }

\def\be{\begin{equation}}
\def\ee{\end{equation}}
\def\bea{\begin{eqnarray}}
\def\eea{\end{eqnarray}}

\newcommand{\Photo}{\includegraphics[height=35mm]{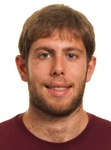}}

\begin{document}
\vspace*{4cm}
\title{Precision measurements of the Lepton-Charge and Forward-Backward Drell-Yan Asymmetries to Enhance the Sensitivity to Broad Resonances of New Gauge Sectors}

\author{ F. Giuli }

\address{CERN, EP Department, CH-1211 Geneva 23, Switzerland}

\maketitle\abstracts{
We study the impact of future measurements of lepton-charge and forward-backward asymmetries in Drell-Yan processes in regions of transverse and invariant masses near the Standard Model gauge bosons peaks to improve the Parton Distribution Functions uncertainties. We study the implications on $W^{'}$ and $Z^{'}$ searches following the reduction of these uncertainties. We find that the sensitivity to the Beyond the Standard Model states is greatly increased with respect to the case of base Parton Distribution Functions sets, thereby enabling one to set more stringent limits on (or indeed discover) such new particles.}

\section{Introduction}
Parton Distribution Functions (PDFs) represent one of the main sources of theoretical systematic uncertainties in hadronic collisions. They influence the potential of experimental searches for discovering or setting exclusion bounds on heavy Beyond the Standard Model (BSM) vector bosons, affecting the experimental sensitivity to test various BSM scenarios. As shown in Ref.~[1] new approaches have been proposed to improve valence quarks PDFs~[2,3]. They are based on combining high-precision measurements of Neutral Current (NC) and Charged Current (CC) Drell-Yan (DY) asymmetries in the mass region close to the $W, Z$ poles. It has been demonstrated that the NC forward-backward asymmetry \AFB is sensitive to the charged-weighted linear combination $(2/3)u_{V}+(1/3)d_{V}$ of up- and down-quark distributions~[4], while the CC lepton-charge asymmetry \AW is sensitive to the difference $u_{V}-d_{V}$.\\
The complementary constraints provided by these two quantities on linearly independent combinations of $u_{V}$ and $d_{V}$ quarks PDFs has been illustrated by a quantitative ``profiling'' using the \texttt{xFitter} fitting framework~[5]. Two different projected luminosity scenarios have been thoroughly examined, namely for the LHC Run 3 (300 fb$^{-1}$) and the High-Luminosity LHC (HL-LHC, 3000 fb$^{-1}$), and it has been demonstrated how the constraints form the \AFB and \AW combination can improve the relative uncertainties on PDF by $\sim$ 20\% in the region of invariant/transverse mass spectra between 2 and 6 TeV, where evidence of $Z^{'}$ and $W^{'}$ states with large widths could be observed.\\
This proceeding summarises the results of Ref.~[6], where the impact of improved PDFs using the afore-mentioned method on the experimental sensitivity of forthcoming $W^{'}$ and $Z^{'}$ searches at the LHC has been analysed, focusing on scenarios characterised by multiple $W^{'}$ and $Z^{'}$ broad resonances and by interference effects of the heavy bosons with each other and with SM gauge bosons in the CC and NC channels.

\section{The 4-Dimensional Composite Higgs Model}
The 4-Dimensional Composite Higgs Model (4DCHM) realisation~[7] of the minimal composite Higgs model illustrated in Ref.~[8] is characterized by the following two parameters: the coupling of the new resonances $g_{\rho}$ and the compositeness scale $f$, with the resonance mass scale being $M\sim fg_{\rho}$~[9]. Three charged $W^{'}$ bosons ($W_{j}$ with $j=1, ..., 3$) and five neutral $Z^{'}$ bosons ($Z_{i}$ with $i=1, ..., 5$) are contained in the 4DCHM scenario. Among these, the $W_{1}$ and $Z_{1}, Z_{4}$ do not couple to leptons and light-quarks, while $W_{3}$ and $Z_{5}$ are too heavy to give non-negligible contributions.\\
This study exploits the strong interference effects between the BSM resonances themselves and between the BSM and SM states, which are responsible for the presence of a statistically significant depletion of events that appears below the Breit-Wigner (Jacobian) peaks in the invariant (transverse) mass distribution in the NC (CC) process. As already demonstrated in Refs.~[10,11], the significance of the depletion of events in a a manner similar to that for the excess of events of the peaks can be defined, and it can be used to extract model-dependent exclusion/discovery limits in the parameter space of a given model. In the following section, such limits from the analysis of both the peak or the dip are presented, for the two specific 4DCHM benchmarks described in Table~\ref{tab:AB_bench}. All these results are based on the assumption that the width of the neutral and charged resonances is fixed to $\Gamma_{X}/M_{X} = 20\%$ (with $X=Z^{'}, W^{'}$).
\begin{table}[t!]
\begin{center}
\begin{tabular}{|c||c|c|c|c|c|c|}
\hline
Benchmark & $f$ [TeV] & $g_\rho$ & $M_{Z_2}$ [TeV] & $M_{Z_3}$ [TeV] & $M_{W_2}$ [TeV] & $M_{W_3}$ [TeV]\\
\hline
A & 3.9 & 1.2 & 5.16 & 5.56 & 5.56 & 6.62\\
\hline
B & 1.5 & 2.2 & 3.39 & 3.45 & 3.45 & 4.67\\
\hline
\end{tabular}
\end{center}
\caption{Parameters for the benchmarks A and B and their heavy gauge bosons masses.}
\label{tab:AB_bench}
\end{table}

\section{Results in the neutral gauge sector}
\begin{figure}[t!]
\begin{center}
\includegraphics[width=0.4\textwidth]{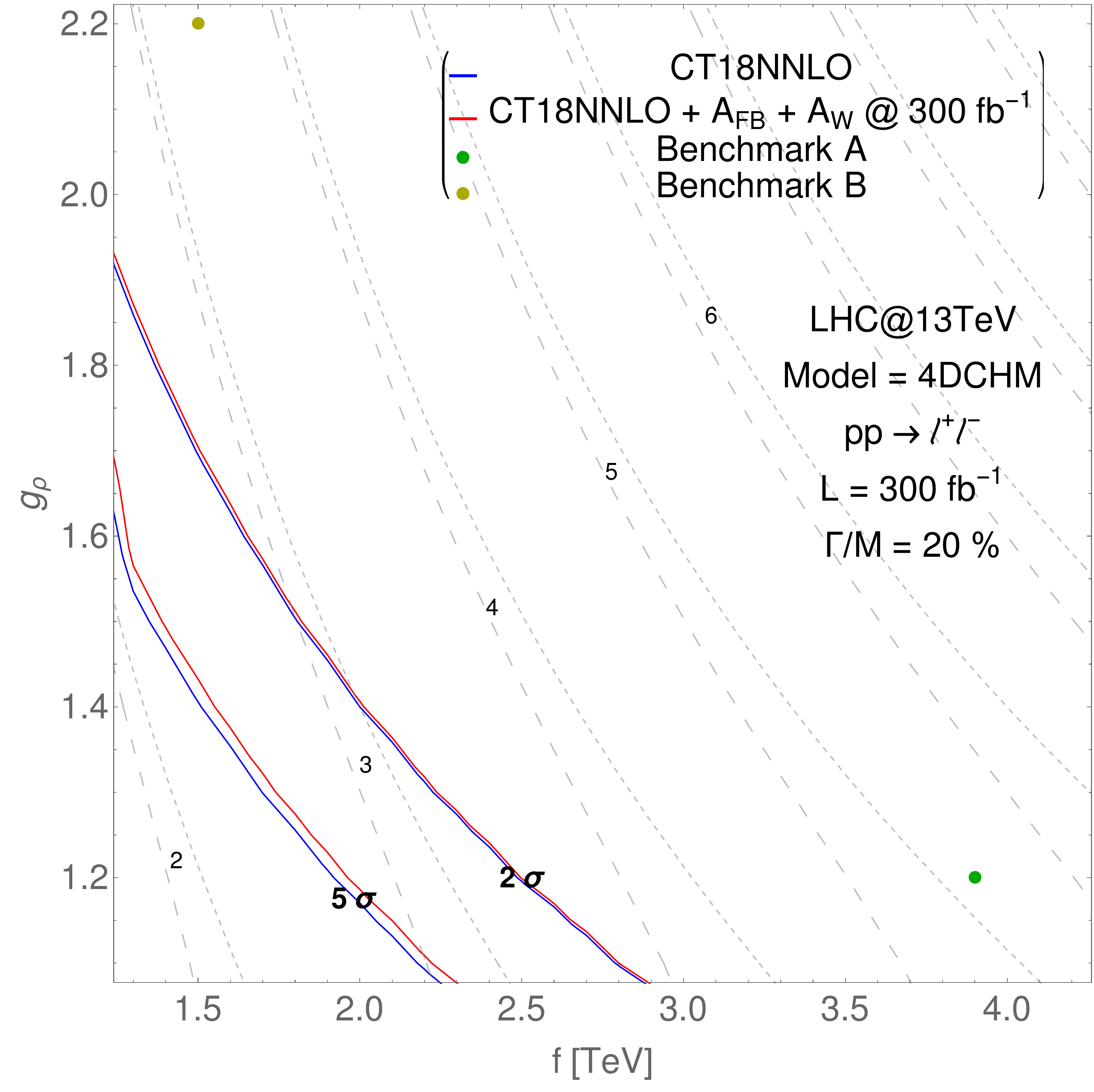}
\includegraphics[width=0.4\textwidth]{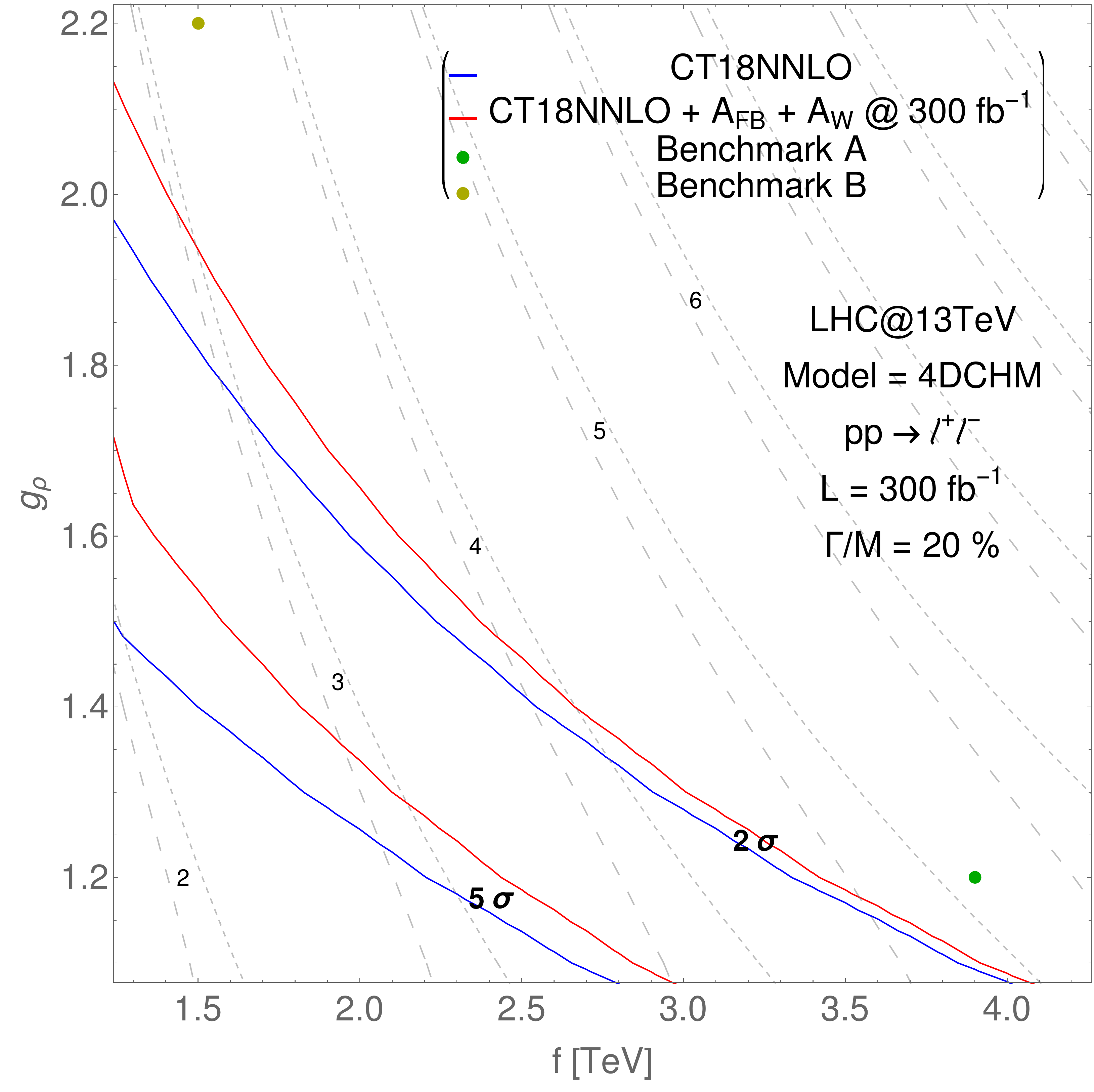}
\end{center}
\caption{Exclusion and discovery limits at 300 fb$^{-1}$ for the peak (left) and for the dip (right) for $Z^{\prime}$ resonances with $\Gamma_{Z^\prime} / M_{Z^\prime} =$ 20\%.
The short (long) dashed contours give the boson mass $M_{Z_2}$ ($M_{Z_3} \simeq M_{W_2}$) in TeV.}
\label{fig:Contour_20_300}
\end{figure}
\begin{figure}[t!]
\begin{center}
\includegraphics[width=0.4\textwidth]{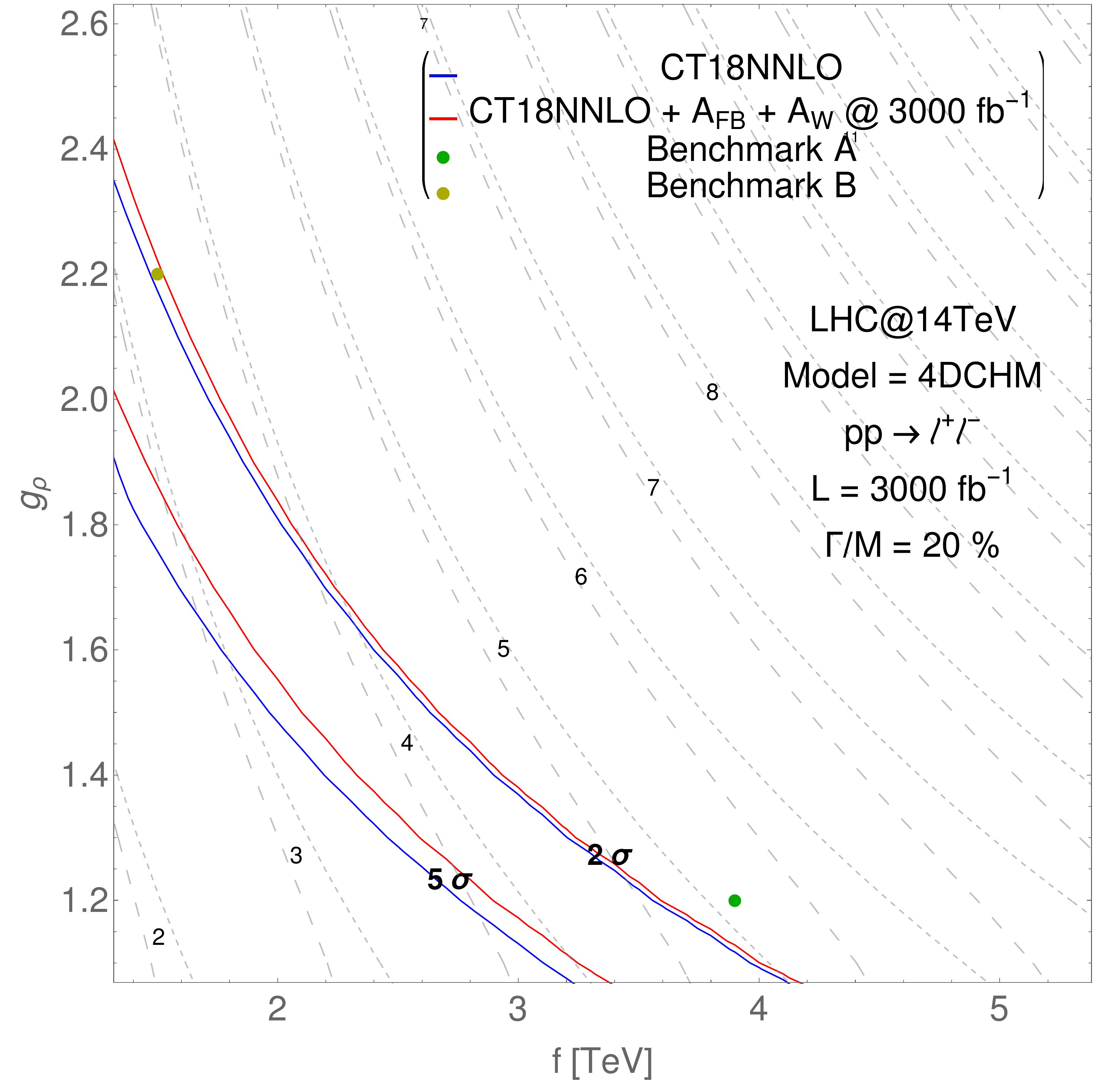}
\includegraphics[width=0.4\textwidth]{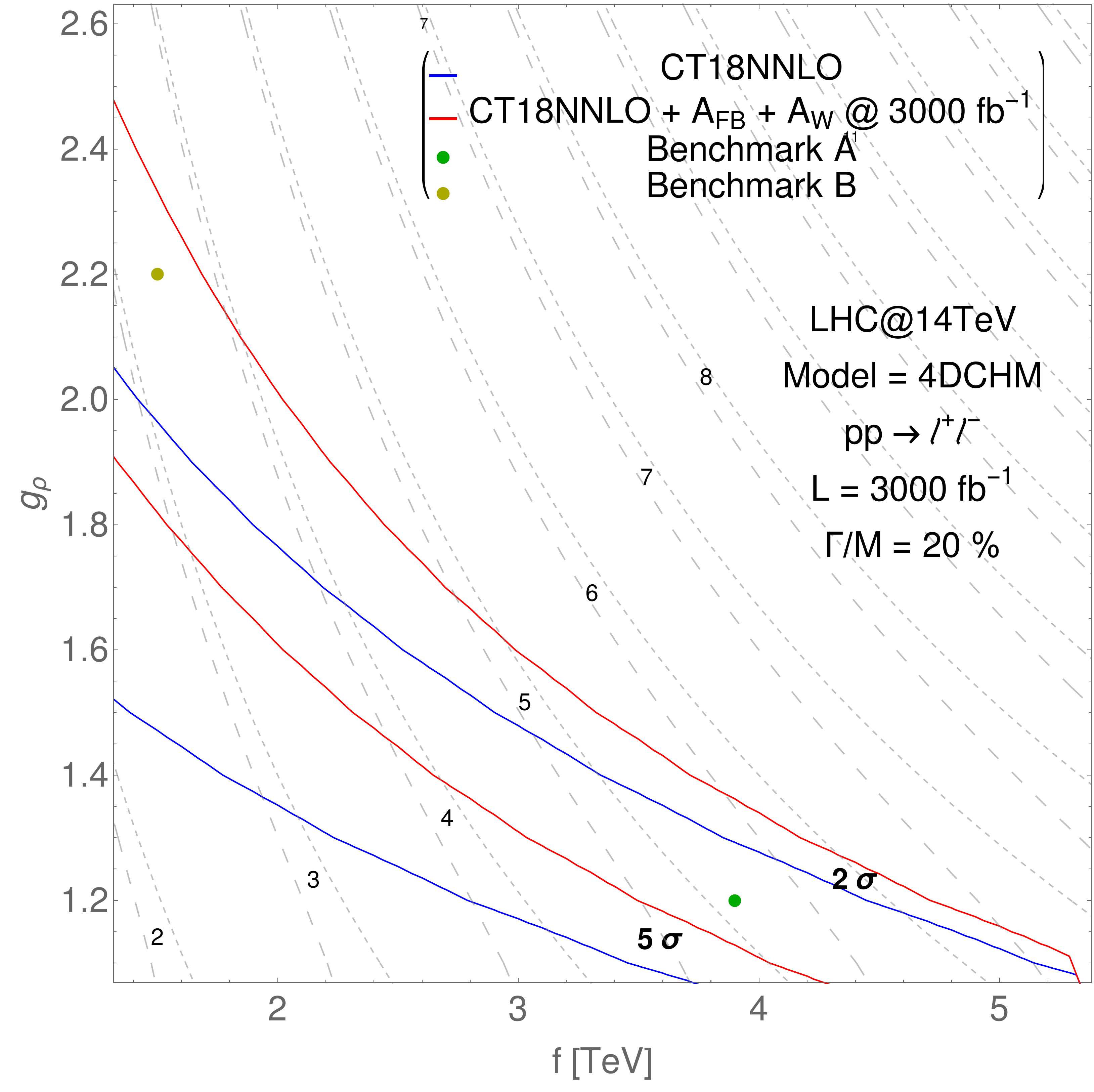}
\end{center}
\caption{Exclusion and discovery limits at 3000 fb$^{-1}$ for the peak (left) and for the dip (right) for $Z^{\prime}$ resonances with $\Gamma / M$ = 20\%.
The short (long) dashed contours give the boson mass $M_{Z_2}$ ($M_{Z_3} \simeq M_{W_2}$) in TeV.}
\label{fig:Contour_20_3000}
\end{figure}

The limits on the model parameter space for the LHC Run 3 with centre-of-mass energy $\sqrt{s} = 13$~TeV and 300 fb$^{-1}$ of integrated luminosity are shown in Figure~\ref{fig:Contour_20_300}. The results for both the peak and the dip in the multi-resonant profile are illustrated, on the left- and right-hand side respectively. The contour plots for the masses of the gauges bosons, $M_{Z_{2}}$ (short-dashed curves) and $M_{Z_{3}}\simeq M_{W_{2}}$ (long-dashed curves) are also given. It can be seen how the sensitivity to the dip is much higher than the one from the peak in the region characterized by large $f$ and small $g_{\rho}$. This is related to the fact that $Z_{2}$ and $Z_{3}$ are very close in mass for small $f$ values, thus  the interference effects accumulate in a confined region below the peak, resulting in a narrow, well pronounced dip. Instead, at large $f$, these BSM resonances are well separated in mass and the interference effects are spread over a wider invariant mass interval, resulting in a less pronounced and broader dip. As expected, the amelioration of the PDF error due to the profiled PDFs is critical in the former model's parameter region and marginal in the latter, with the overall result that the dip selection is vastly more constraining than the peak one.\\
Similarly, Figure~\ref{fig:Contour_20_3000} shows the limits on the model parameters space for the HL-LHC stage with an integrated luminosity of 3000 fb$^{-1}$ and $\sqrt{s} = 14$~TeV. In this setup, the peak of benchmark A would still be below the experimental sensitivity, while the peak of benchmark B, if the improved PDFs are employed, would be right below the 2$\sigma$ exclusion. When exploiting the depletion of events in the dip below the peak, the sensitivity on the model increases remarkably. Furthermore, the improvement on the PDF also has a very large impact particularly in the region of small $f$ and large $g_{\rho}$. Taking into account the reduction of PDF uncertainty, benchmark A would be now at the edge of the 5$\sigma$ discovery, while the sensitivity on benchmark B would almost reach 3$\sigma$.\\
Moreover, we explored the potential for evidence or discovery at the HL-LHC. It has been found that while the improvement in significance due to the profiles PDFs is more marginal for the peak region (between 3\% and 10\%), it becomes sizeable in the dip case, as here significances grow by an amount from 40\% to 90\% for the benchmarks A and B, respectively.

\section{Results in the charged gauge sector}
\begin{figure}[t!]
\begin{center}
\includegraphics[width=0.4\textwidth]{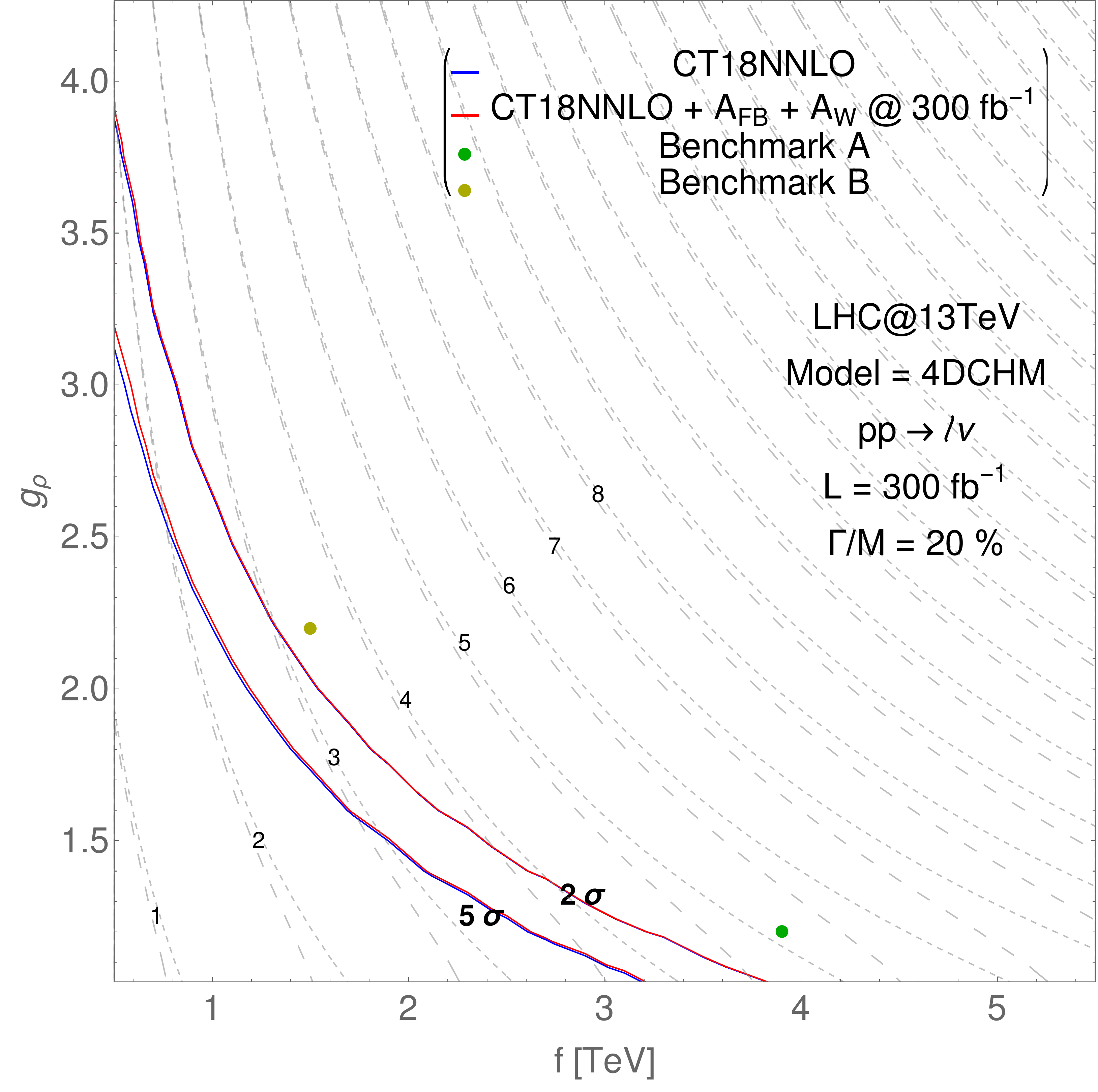}
\includegraphics[width=0.4\textwidth]{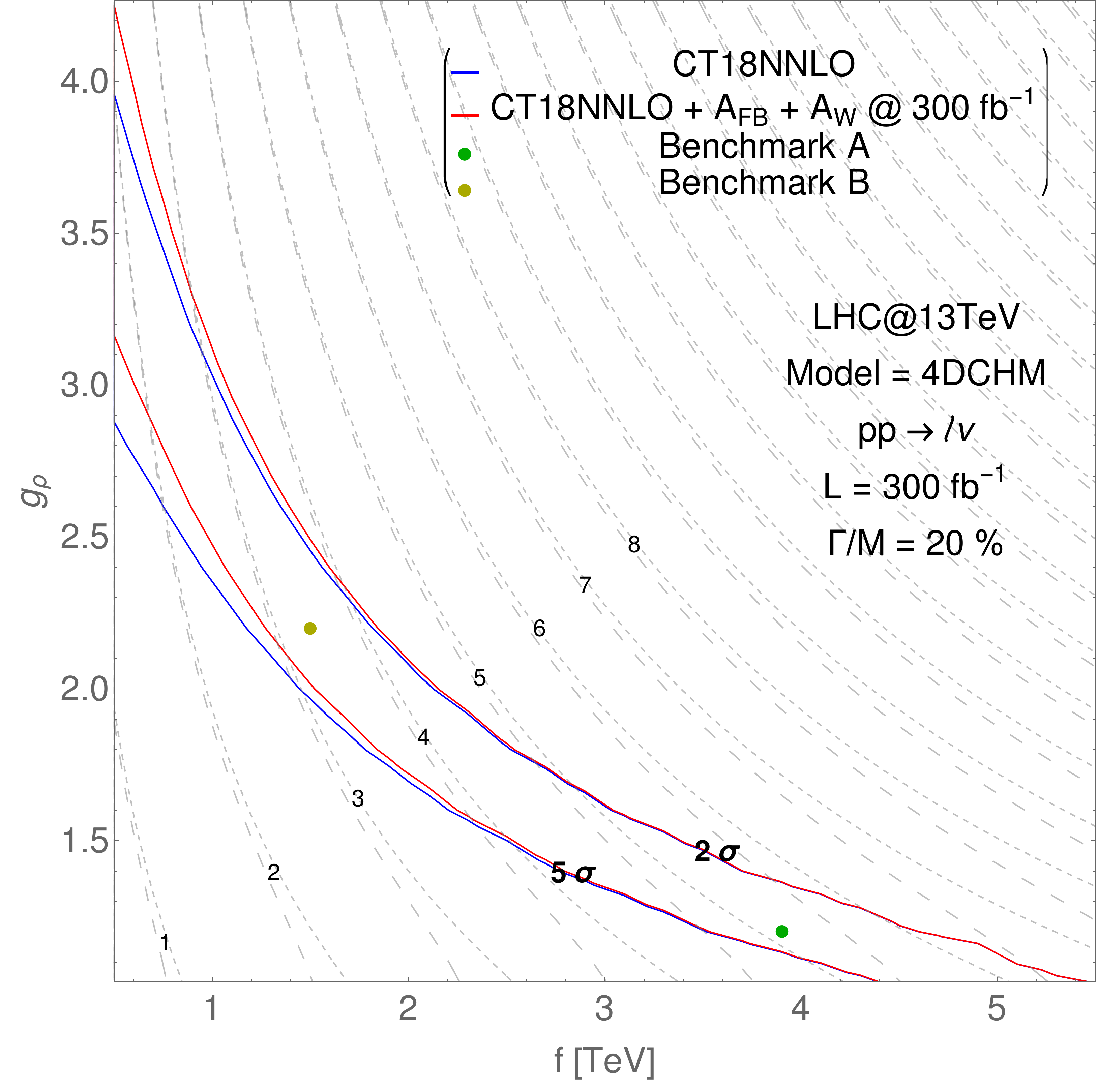}
\end{center}
\caption{Exclusion and discovery limits at 300 fb$^{-1}$ for the peak (left) and for the dip (right) for $W^{\prime}$ resonances with $\Gamma / M$ = 20\%.
The short (long) dashed contours give the boson mass $M_{Z_2}$ ($M_{Z_3} \simeq M_{W_2}$) in TeV.}
\label{fig:Contour_Wpr_20_300}
\end{figure}
\begin{figure}[t!]
\begin{center}
\includegraphics[width=0.4\textwidth]{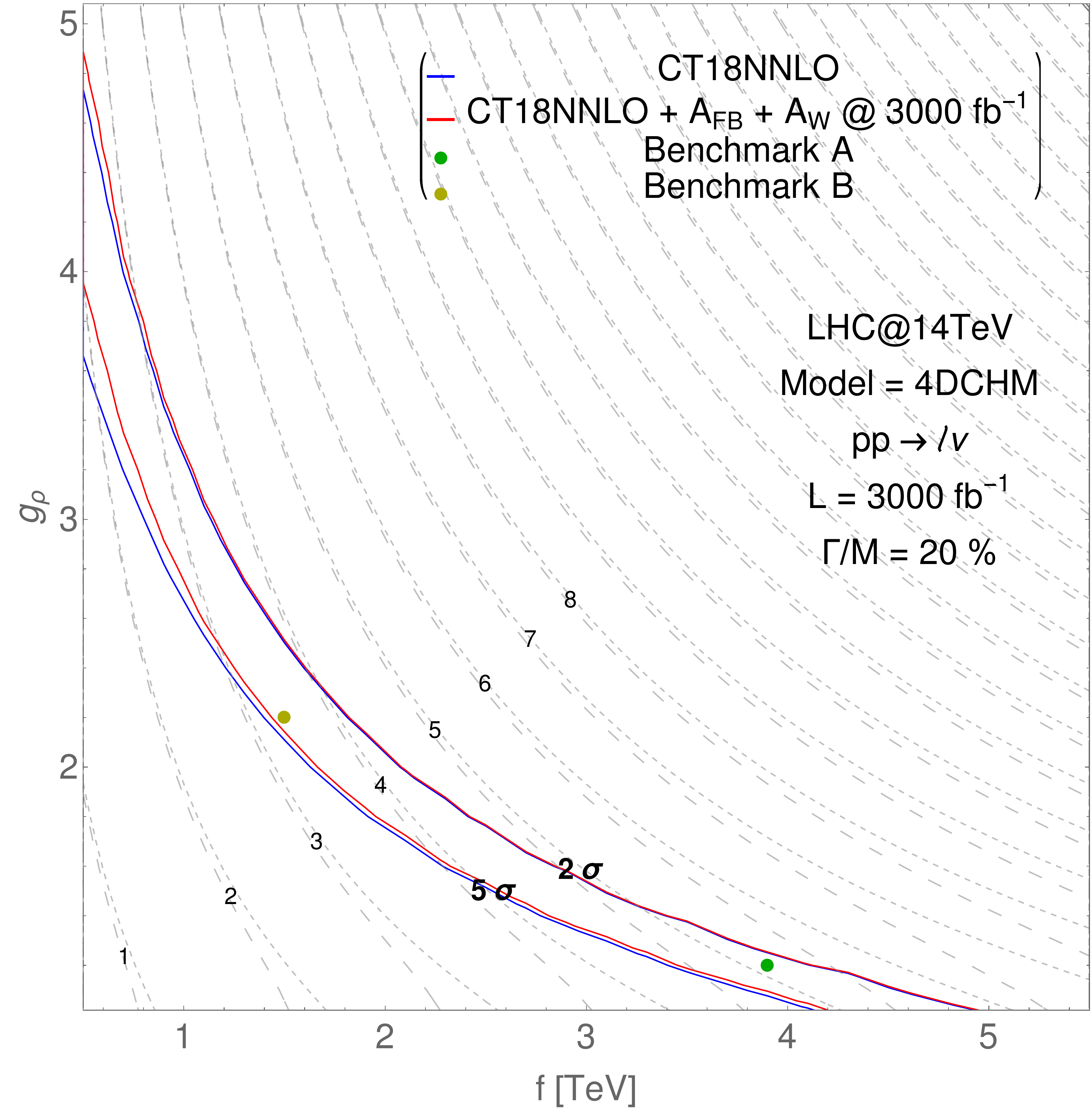}
\includegraphics[width=0.4\textwidth]{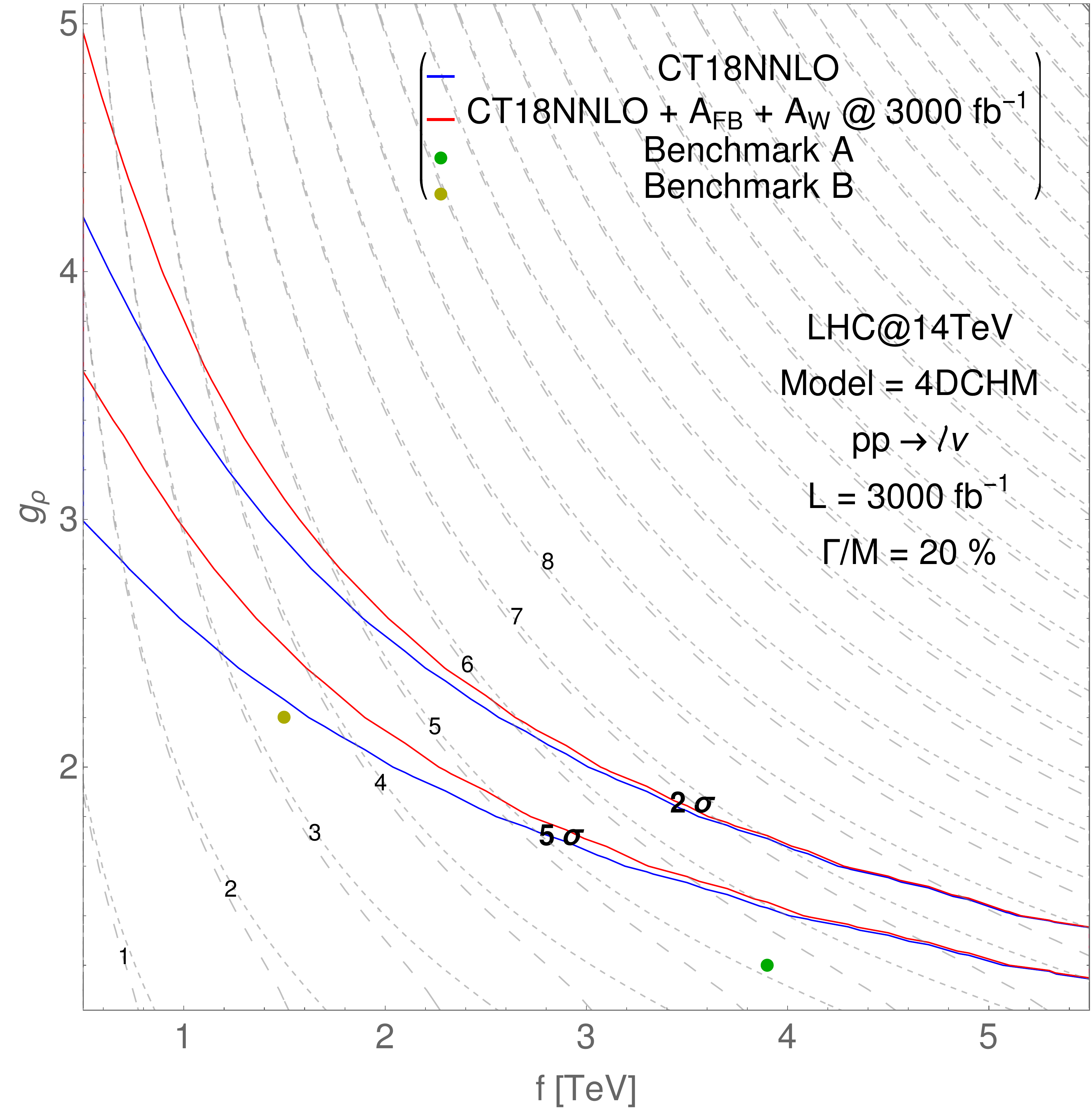}
\end{center}
\caption{Exclusion and discovery limits at 3000 fb$^{-1}$ for the peak (left) and for the dip (right) for $W^{\prime}$ resonances with $\Gamma / M$ = 20\%.
The short (long) dashed contours give the boson mass $M_{Z_2}$ ($M_{Z_3} \simeq M_{W_2}$) in TeV.}
\label{fig:Contour_Wpr_20_3000}
\end{figure}
Figure~\ref{fig:Contour_Wpr_20_300} shows the limits on the model parameter space for the LHC Run 3 with $\sqrt{s} = 13$~TeV and an integrated luminosity of 300 fb$^{-1}$. It is clear from the plots that the CC channel provides more stringent limits on the 4DCHM, as the production cross section if higher for charged gauge bosons than for the neutral ones. However, since we are dealing with a single boson resonance, the interference effects which generate the dip are now weaker. The sensitivity to the dip significantly exceeds that of the peak only in the large-$f$ region, with small improvement due to the reduction of PDF uncertainty. In this setup, the peaks of the two benchmarks still remain below 2$\sigma$. In contrast, exploiting the depletion of events below the Jacobian peaks, both benchmarks would reach a 3$\sigma$ significance.\\
The limits on the model parameter for the HL-LHC stage with an integrated luminosity of 3000 fb$^{-1}$ and $\sqrt{s} = 14$~TeV are shown in Figure~\ref{fig:Contour_Wpr_20_3000}. We note a significant improvement in the sensitivity to the model from the analysis of the dip in comparison with the analysis of the peak for the region with large values of the compositeness scale $f$. In the complementary region, the signal of the peak generally has a larger significance, particularly for discovery purposes. The analysis of the dip, however, can be competitive in drawing exclusions, especially when the profiled PDFs are employed. In this setup, the two benchmarks A and B are both within HL-LHC reach, particularly so in the presence of profiled PDFs.\\
As for the neutral gauge sector, the HL-LHC potential to test the possibility of the existence of a $W^{'}$ state of the 4DCHM sector has been studied. In particular, for the charged gauge sector, we see that the sensitivity to the model is greatly increased, as the peak of benchmarks A and B reach about 3$\sigma$ and 4$\sigma$ deviations. Indeed, the sensitivity to the model from the depletion of events driven by interference effects is predominant in this context , as both benchmarks give deviations $> 5\sigma$ in the low mass tail region. As visible from the right plot of Figure~\ref{fig:Contour_Wpr_20_3000}, the improvement in the PDFs would benefit the experimental analysis especially in the parameter space region with large $g_{\rho}$ and small $f$.

\section{Summary and outlook}
In this proceeding it has been shown that improving the non-perturbative Quantum Chromodynamics (QCD) systematic uncertainties associated with the initial state PDFs enhances significantly the potential of forthcoming LHC experiments in both discovering and setting exclusion limits on broad $Z^{'}$ and $W^{'}$ resonances. The possibility of exploiting model-dependent effects due to the interference of heavy BSM gauge bosons with the SM has been explored, in order to extend the experimental sensitivity of dedicated searches. In particular, in a well-defined theoretical framework like the 4DCHM, it has been found that the analysis of the dip often provides more stringent limits than the signal coming from the peak, with the reduction of systematic PDF errors playing a crucial part in this conclusion.\\
Finally, it can be emphasized that results obtained in $W^{'}$ searches via the CC channel can be potentially used to constrain $Z^{'}$ properties as well, in a manner more stringent than that afforded by direct searches for it in the NC channel.~\footnote{Phenomenological studies based on combined NC and CC searches have been performed in Ref.~[12] in the context of the Sequential Standard Model and Left-Right Symmetric Models.} In particular, we stress that this is a remarkably efficient approach in the case of profiled PDFs, since up to a 1 TeV more sensitivity can indirectly be gained to $Z^{'}$ masses in $W^{'}$ direct searches with respect to the scope of $Z^{'}$ direct searches themselves. We further remark that while these conclusions have general validity, the precise size of the quantitative improvements from the combination of NC and CC analyses, as well as from the use of profiled PDFs, depends on the specific realisation of the BSM resonances and in particular on their widths.

\section*{Acknowledgments}
The results presented in this article are obtained in collaboration with J. Fiaschi, F. Hautmann and S. Moretti.

\end{document}